\documentstyle[12pt,epsf]{article}

\def\gtap{\raisebox{-.4ex}{\rlap{$\sim$}} \raisebox{.4ex}{$>$}}
\def\ltap{\raisebox{-.4ex}{\rlap{$\sim$}} \raisebox{.4ex}{$<$}}

\newcommand{\agt}{\mathrel{\raisebox{-.6ex}{$\stackrel{\textstyle>}{\sim}$}}}

\begin{document}
\font\fortssbx=cmssbx10 scaled \magstep2
\hbox to \hsize{
\hskip.5in \raise.1in\hbox{\fortssbx University of Wisconsin - Madison}
\hfill\vbox{\hbox{\bf MAD/PH/828}
            \hbox{April 1994}} }

\vglue1.5cm

\begin{center}
{\bf  Supergravity Solutions in the Low-$\tan\beta$ $ \lambda_t$
Fixed Point Region}
\footnote{%
Talk presented by P. Ohmann at the {\it Second IFT Workshop: Yukawa
Couplings and the Origins of Mass}, Gainesville, Florida, February
11--13, 1994.}\\[.5cm]
{\small V.~Barger, M.~S.~Berger, P.~Ohmann}\\[.2cm]
{\small\it Physics Department, University of Wisconsin, Madison, WI 53706, USA}
\end{center}

\vglue.6cm

\abstract{
There has been much discussion in the literature about applying
the radiative
electroweak symmetry breaking (EWSB) requirement to GUT models with
supergravity. We motivate and discuss the application of the
EWSB requirement to the low $\tan\beta$ fixed-point region
and describe the solutions we find.}

\section{Introduction}

Improvements in LEP data over the past few years have generated
significant excitement at the prospect of grand unification within
the Minimal Supersymmetric Standard Model (MSSM) \cite{susygut}.
In addition to the gauge coupling unification suggested by LEP,
Yukawa unification -- in particular $\lambda_b(M_G) = \lambda_{\tau}(M_G)$
\cite{ceg} -- has been extensively studied, both at the one-loop and
two-loop levels\cite{bbo}. Such a constraint places significant
restrictions on the allowed parameter space, especially that of $m_t$
and $\tan\beta$. For values of $m_b(m_b)$ within the
range $4.25 \pm 0.10$ GeV \cite{gl}, the resulting allowed parameter space
lies almost exclusively within the fixed - point region, as defined
by $\lambda _i^G\agt 1$ for $i = t, b$, and/or ${\tau}$
\cite{bbo},\cite{bbop}--\cite{nmssm}.

If one makes only the additional assumption that $m_t(m_t) \ltap 175$ GeV
(consistent with the recently released CDF measurement
$m^{pole}_t = 174 \pm 10 {\scriptsize \begin{array}{c}+13 \\ -12 \end{array}}$
GeV\cite{cdf1} which corresponds to a running mass
$m_t(m_t) \simeq 166 \pm 10 \pm 13$
GeV,
then one is restricted to two very narrow regions in the $m_t$,
$\tan \beta$ plane. One of these regions,
the low $\tan \beta$ fixed-point region, has been the
focus of our recent renormalization group analysis
with supersymmetric grand unification\cite{bbop},
and we therefore
examine whether these solutions satisfy
the additional constraint imposed by Radiative Electroweak
Symmetry Breaking (EWSB).

\section{Fixed Points and $\lambda_b = \lambda_{\tau}$ Unification}

Fixed-points arise naturally from imposing
$\lambda_b = \lambda_{\tau}$ unification at the GUT scale
along with the typically allowed range for the
bottom quark mass $4.25 \pm 0.10$ GeV.
Figure 1 shows the allowed parameter space for
$\lambda_b = \lambda_{\tau}$ unification, and Figure 2
shows contours of Yukawa couplings (at $M_{GUT}$).

\begin{center}
\epsfxsize=5.75in
\hspace*{0in}
\epsffile{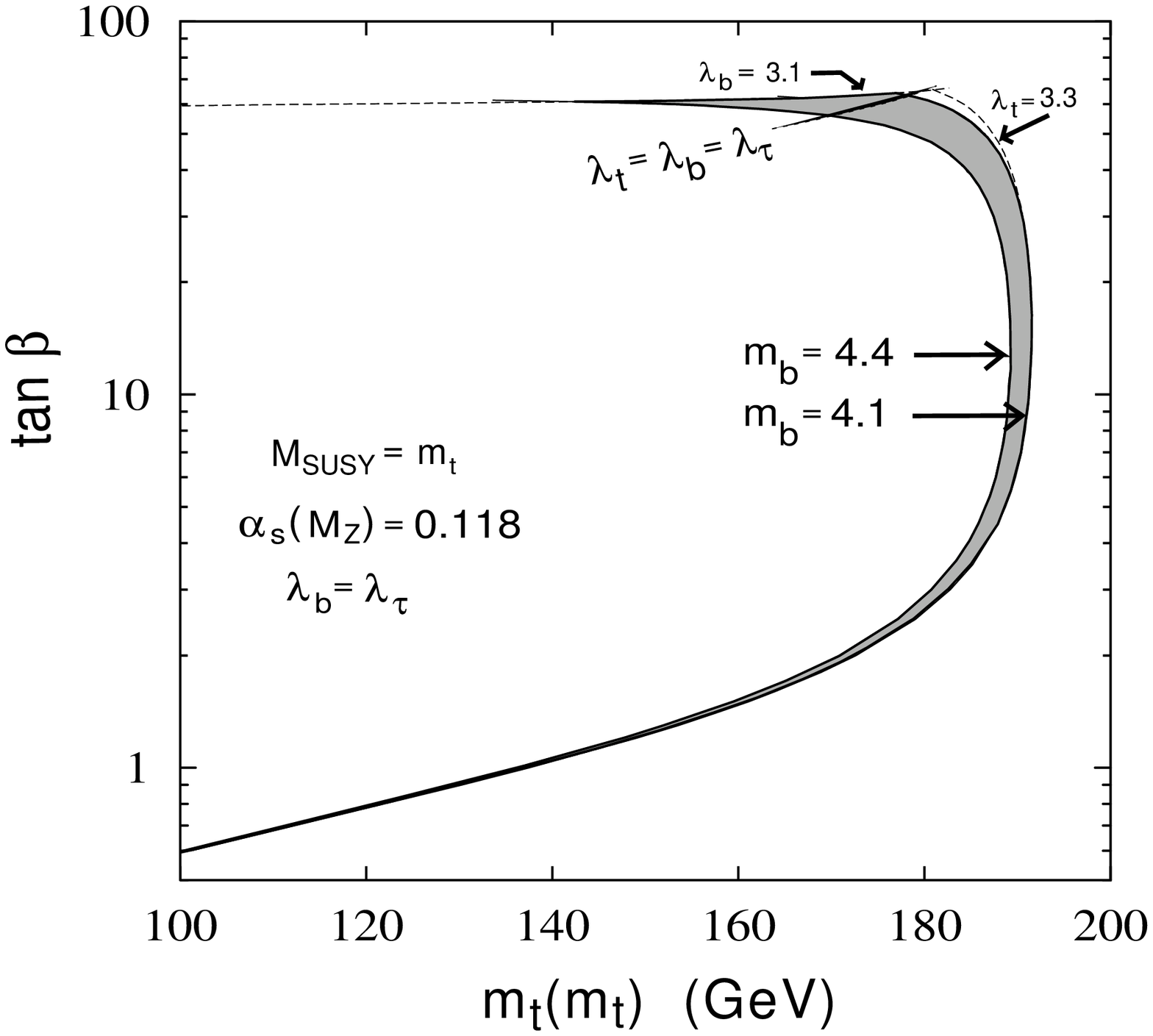}

\smallskip
\parbox{5.5in}{\small Fig.~1. Contours of constant $m_b(m_b)$ in the
$m_t(m_t),\tan\beta$ plane (from Ref.~\cite{bbo}).}
\end{center}

\begin{center}
\epsfxsize=4in
\hspace*{0in}
\epsffile{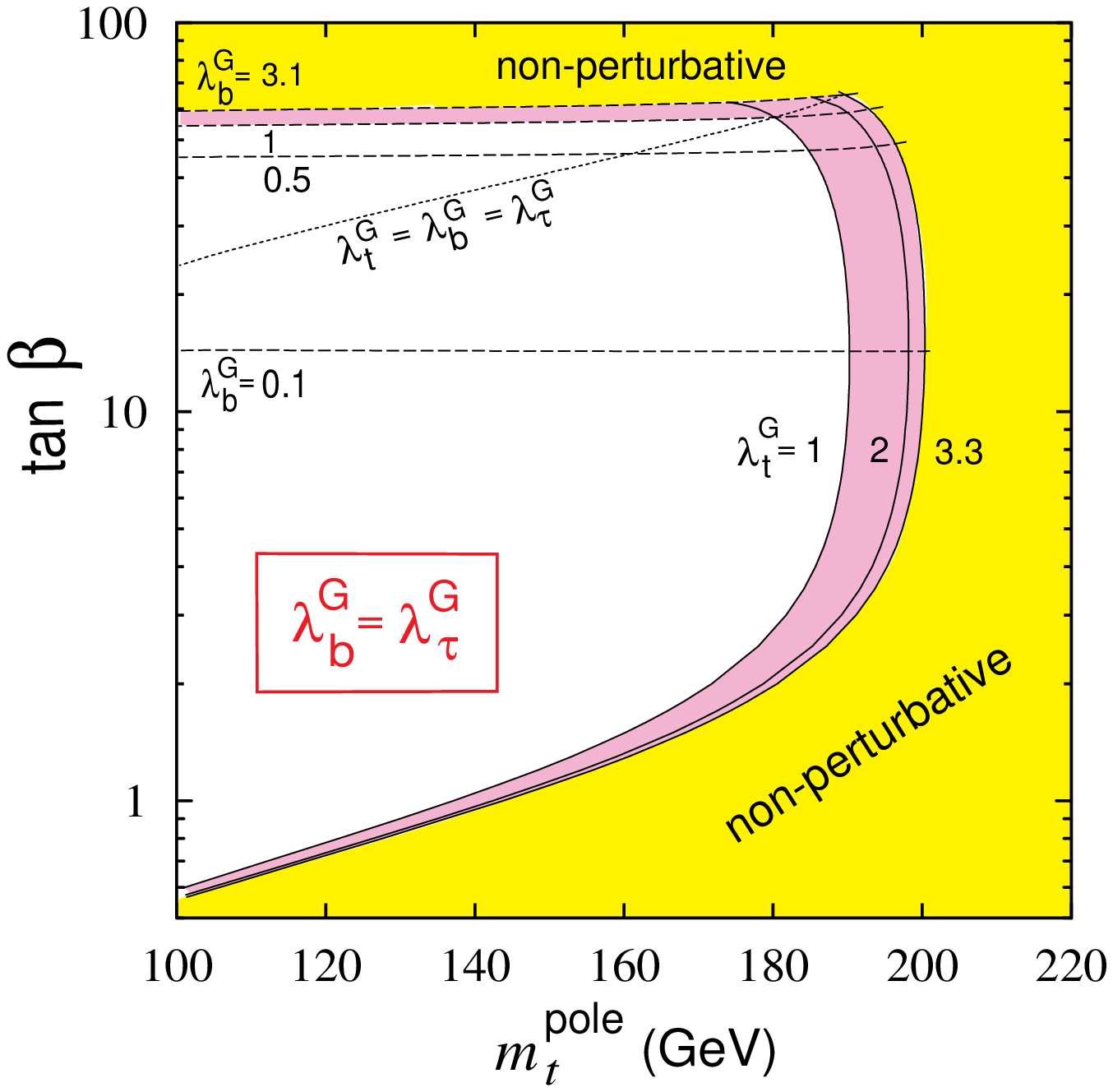}

\medskip
\parbox{5.5in}{\small Fig.~2. The fixed-point regions are given by Yukawa
couplings at the GUT scale being larger than about 1 ($\lambda _i^G\agt 1$).
Even larger values of the Yukawa couplings results in a breakdown of
perturbation theory.}
\end{center}
Note that Figure 1 is a subset of Figure 2 (allowing for
the small $\sim$ 5-10 GeV difference between $m_t(m_t)$ and $m^{pole}_t$);
in fact, imposing this $m_b$ mass constraint ensures the
fixed-point nature of the solutions. Figure 3 shows the typical
evolution of $\lambda_t$ for these solutions.

\begin{center}
\epsfxsize=4.0in
\hspace{0in}
\epsffile{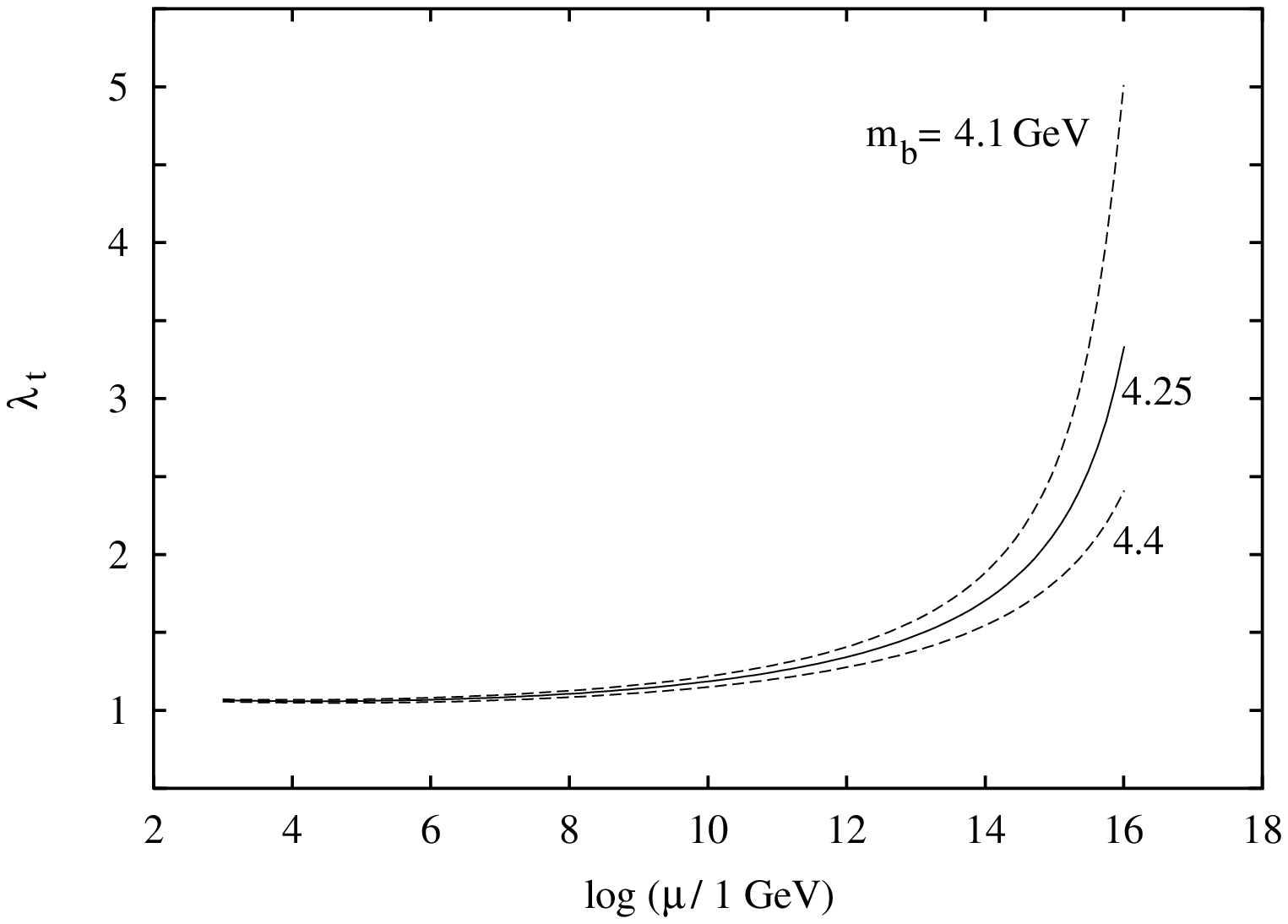}

\smallskip
\parbox{5.5in}{\small Fig.~3. If $\lambda _t$ is large at $M_G^{}$, then
the renormalization group equation causes $\lambda _t(Q)$ to evolve rapidly
towards an infrared fixed point as $Q \rightarrow m_t$ (from Ref.~\cite{bbo}).}
\end{center}

\smallskip

As described by Figures 1 and 2,
the allowed $m_t$-$\tan\beta$ parameter space
can be divided into three distinct regions:

   1) $\tan\beta \ltap 2$ ($\lambda_t$ fixed point)

   2) $\tan\beta \gtap 50$ ($\lambda_b = \lambda_{\tau}$ fixed point)

   3) $2 \ltap \tan\beta \ltap 50$ ($\lambda_t$ fixed point)

It should be noted that threshold corrections, if large,  may either
enforce or mitigate
the fixed-point nature of some of the solutions
\cite{cpw}--\cite{lp},\cite{bh}--\cite{wright}.
If $m_t \ltap 175$ GeV,
then only the first two regions remain. While both solution sets
may still be viable, there are criteria which seem
to favor the first region:
namely, the large $\tan \beta $ region typically results
in large threshold corrections
and in large enhancements to flavor changing neutral currents in processes
like $b\to s\gamma$ and $B\overline{B}$ mixing and to proton
decay\cite{an}--\cite{hmy}. However, it is possible these may
be successfully eliminated
by assuming certain symmetries\cite{hrs}.

One of the most important aspects of the
large top mass is that it makes possible an
understanding of the radiative breaking of
the electroweak symmetry; the large
top quark Yukawa drives a Higgs mass-squared negative.
However, when both the top
and bottom quark Yukawas are the same size at the GUT scale, one must rely
on the difference in their hypercharges
to effect the symmetry breakdown \cite{copw2}.

The low $\tan\beta$ $\lambda_t$ fixed-point region can be well described
by the following
relation between the top quark mass
and $\tan \beta $.
\begin{eqnarray}
\lambda_t(m_t) &=& {\sqrt2 m_t(m_t)\over v\sin\beta}
\approx 1.1 \ \Rightarrow
\ m_t(m_t)\approx {v\over {\sqrt{2}}}\sin \beta=(192 {\rm GeV})
\sin \beta \;
\end{eqnarray}
Converting this relation to the top quark pole mass yields\cite{bbo,bbop}
\begin{equation}
m_t^{\rm pole }\approx(200 {\rm GeV})\sin \beta \;.
\end{equation}

\section{Electroweak Symmetry Breaking}

There has been much discussion in the literature about imposing
a Radiative Electroweak Symmetry Breaking (EWSB) constraint
on GUT models \cite{nac}--\cite{kkrw}; in particular we address this issue
with regard to
the low $\tan\beta$ $\lambda_t$ fixed-point region.

The EWSB constraint is enforced by minimizing the effective
Higgs potential; at tree-level this is given by:
\begin{eqnarray}
V_0&=&(m_{H_1}^2+\mu ^2)|H_1|^2+(m_{H_2}^2+\mu
^2)|H_2|^2+m_3^2(\epsilon_{ij}{H_1}^i{H_2}^j+{\rm h.c.})
\nonumber \\
&&+{1\over 8}(g^2+g^{\prime 2})\left [|H_1|^2-|H_2|^2\right ]^2
+{1\over 2}g^2|H_1^{i*}H_2^i|^2\;, \label{tree}
\end{eqnarray}
where $m^{2}_{H_1}$, $m^{2}_{H_2}$, and
$m^{2}_3 = B\mu$ are soft-supersymmetry breaking parameters,
$\epsilon _{ij}$ is the
antisymmetric tensor, and $H_1$ and $H_2$ are complex
doublets given by
\begin{eqnarray}
H_1&=& \left( \begin{array}{c}
{{1\over{\sqrt{2}}}(\psi_1 + v_1 + i\phi_1)} \\
H^{-}_1
\end{array} \right)\;,
\nonumber \\
H_2&=&\left( \begin{array}{c}
H^{+}_2 \\
{{1\over{\sqrt{2}}}(\psi_2 + v_2 + i\phi_2)}
\end{array} \right)\;.
\nonumber \\
\end{eqnarray}
Minimizing this potential with respect to the two real
components of the neutral Higgs fields
$\psi_1$ and $\psi_2$ yields the tree-level EWSB minimization
conditions:

\begin{eqnarray}
{1\over 2}M_Z^2&=&{{m_{H_1}^2-m_{H_2}^2\tan ^2\beta }
\over {\tan ^2\beta -1}}-\mu ^2 \;,  \label{treemin1} \\
-B\mu &=&{1\over 2}(m_{H_1}^2+m_{H_2}^2+2\mu ^2)\sin 2\beta \;.
\label{treemin2}
\end{eqnarray}
The masses in these equations are running masses that depend on the scale
$Q$ in the RGEs that describe their evolution. Hence the solutions
obtained are functions of the scale $Q$.
Equations~(\ref{treemin1}) and (\ref{treemin2}) are particularly convenient
since the gauge couplings dependence (the D-terms
in the language of supersymmetry) is isolated in Eq.~(\ref{treemin1}).
In addition, these minimization equations are readily solvable
(even at the one-loop level) with the ambidextrous approach, which
we describe in the next section.

The minimization equations
also clearly show the fine-tuning problem that may be present in
the radiative breaking of the electroweak symmetry. For large values of
$|\mu |$, there must be a cancellation between large terms on the right hand
side of equation~(\ref{treemin1})
to obtain the correct experimentally measured $M_Z^{}$ (or equivalently
the electroweak scale). For $\tan \beta $ near one, a cancellation of
large terms must occur.

A heavy top quark produces large corrections to the Higgs potential of the
MSSM\cite{msb}.
Gamberini, Ridolfi, and Zwirner showed\cite{grz} that the tree-level
Higgs potential is inadequate for the purpose of analyzing radiative breaking
of the electroweak symmetry because the tree-level Higgs vacuum expectation
values
$v_1$ and $v_2$ are very sensitive to the scale at which the renormalization
group equations are evaluated.
The one-loop contribution to the effective potential is given by
\begin{eqnarray}
\Delta V_1&=&{1\over {64\pi ^2}}{\rm Str}\left [{\cal M}^4
\left (\ln {{{\cal M}^2}\over {Q^2}}-{3\over 2}\right )\right ]\;, \label{V1}
\end{eqnarray}
where $\Delta V_1$ is  given in the dimensional reduction
($\overline{DR}$) renormalization scheme\cite{dimred}.
The supertrace is defined as
${\rm Str} f({\cal M}^2)=\sum _iC_i(-1)^{2s_i}(2s_i+1)f(m_i^2)$ where
$C_i$ is the color degrees of freedom and $s_i$ is the spin of the
$i^{th}$ particle.

The one-loop corrections to the
Higgs potential effectively moderates this sensitivity to the scale $Q$.
The one-loop corrections are conveniently calculated using the tadpole
method\cite{madph801},\cite{tadpole},\cite{sher}.
The one-loop corrected minimization conditions can then be used to generate
a complete supersymmetric particle spectrum which satisfies EWSB.
Including only the leading contribution coming from the top quark loop (and
neglecting the D-term contributions to the squark masses)
one obtains the expressions
\begin{eqnarray}
{1\over 2}M_Z^2&=&{{m_{H_1}^2-m_{H_2}^2\tan ^2\beta }
\over {\tan ^2\beta -1}}-\mu ^2
-{{3g^2m_t^2}\over {32\pi ^2M_W^2\cos 2\beta}}\Biggl[ 2f(m_t^2)
-f(m_{\tilde{t}_{1}}^2)-f(m_{\tilde{t}_{2}}^2) \nonumber\\
&& \hspace{2.5in} {} +{{f(m_{\tilde{t}_{1}}^2)
-f(m_{\tilde{t}_{2}}^2)}
\over {m_{\tilde{t}_1}^2-m_{\tilde{t}_2}^2}}
\Bigl( (\mu \cot \beta)^2-A_t^2\Bigr) \Biggr] \;, \nonumber\\
\label{treemin3}\\
-B\mu &=&{1\over 2}(m_{H_1}^2+m_{H_2}^2+2\mu ^2)\sin 2\beta
-{{3g^2m_t^2\cot \beta }\over {32\pi ^2M_W^2}}\Biggl[ 2f(m_t^2)
-f(m_{\tilde{t}_{1}}^2)-f(m_{\tilde{t}_{2}}^2) \nonumber\\
&& \hspace{1.75in} {} -{{f(m_{\tilde{t}_{1}}^2)-f(m_{\tilde{t}_{2}}^2)}
\over {m_{\tilde{t}_1}^2-m_{\tilde{t}_2}^2}}
(A_t+\mu \cot \beta )(A_t+\mu \tan \beta )\Biggr]\;, \nonumber \\
\label{treemin4}
\end{eqnarray}
where
\begin{eqnarray}
f(m^2)=m^2\left (\ln {{m^2}\over {Q^2}}-1\right )\;.
\end{eqnarray}
The extra one-loop contribution included above renders the solution less
sensitive to the scale $Q$\cite{aspects}--\cite{Ramond},
\cite{madph801},\cite{cc},
as can be shown explicitly by examining the
relevant renormalization group equations for the parameters that enter into
the minimization conditions.
The complete expressions for the one-loop contributions can be found in
Ref.~\cite{madph801}.
The fine-tuning problem is alleviated somewhat, but not entirely, by the
inclusion of one-loop corrections to the Higgs potential.
As our naturalness criterion we require
\begin{eqnarray}
&&|\mu (m_t)|< 500\ {\rm GeV}\;.
\end{eqnarray}

\section{Ambidextrous Approach}
Other RGE studies of the supersymmetric particle spectrum have evolved from
inputs at the GUT scale (the top-down method\cite{hempf2})
or from inputs at the electroweak
scale (the bottom-up approach\cite{op2}).
The ambidextrous approach \cite{Kelley}
incorporates some boundary conditions at both electroweak and
GUT scales.
We specify
$m_t$ and $\tan \beta$ at the electroweak scale (along with $M_Z$ and $M_W$)
and the common gaugino mass $m_{1 \over 2}$,
scalar mass $m_0$, and trilinear coupling $A^G$ at the GUT scale.
The soft supersymmetry breaking parameters are evolved from the GUT scale to
the electroweak scale and then $\mu (M_Z)$ and $B(M_Z)$
(or $\mu (m_t)$ and $B (m_t)$) are determined by
the one-loop minimization equations.

This strategy is effective because the RGEs for the
soft-supersymmetry breaking parameters do not depend on
$\mu $ and $B$. This method has two powerful advantages:
First, any point in the $m_t$ -- $\tan \beta$ plane can be readily
investigated in specific supergravity
models since $m_t$ and $\tan \beta $ are taken as inputs.
Second, the minimization equations

are easy to solve in the ambidextrous approach:
equation (\ref{treemin3}) can be solved iteratively for $\mu(M_Z)$
(to within a sign), and
then equation (\ref{treemin4}) explicitly gives $B(M_Z)$.
We stress the numerical simplicity: no derivatives need be calculated
and no functions need to be numerically minimized.

\section{Low $\tan\beta$ Fixed Point Solutions}

We now describe our numerical approach in more detail.
Starting with our low-energy choices for $m_t$, $\tan \beta$, $\alpha_3$,
and $m_b$ (and using the experimentally determined values for
$\alpha_1$, $\alpha_2$ and $m_\tau$\cite{pdb}), we integrate
the MSSM RGEs from $m_t$ to $M_G$ with $M_G^{}$ taken
to be the scale $Q$ at which  $\alpha_1 (Q)$ =  $\alpha_2 (Q)$.
We then specify $m_{1 \over 2}$, $m_0$, and $A$ at $M_G$, and integrate
back down to $m_t$ where we solve the full one-loop
minimization equations (see Ref.~\cite{madph801})
for $\mu(m_t)$
and $B(m_t)$. We can then integrate the RGEs back to $M_G$
to obtain  $\mu(M_G)$ and $B(M_G)$.

In particular, we choose values of $m_t$, $\tan \beta$, $\alpha_3$,
and $m_b$ representative of the low $\tan \beta$ $\lambda_t$
fixed point region. In addition to requiring EWSB to be satisfied,
we impose the following experimental bounds:

\bigskip

\begin{center}
{\small Table 1.  Approximate experimental bounds.}
\medskip
\begin{tabular}{|c|c|}
\hline
Particle& Experimental Limit (GeV)
\\ \hline \hline
gluino& 120
\\ \hline
squark, slepton& 45
\\ \hline
chargino& 45
\\ \hline
neutralino& 20
\\ \hline
light higgs& 60
\\ \hline
\end{tabular}
\end{center}
Together with our naturalness criteria
$|\mu (m_t)|<500$~GeV, these bounds give the allowed region in the
$m_0,m_{1/2}$ plane shown as the shaded areas in Fig.~4.

\begin{center}
\epsfxsize=4in
\hspace*{0in}
\epsffile{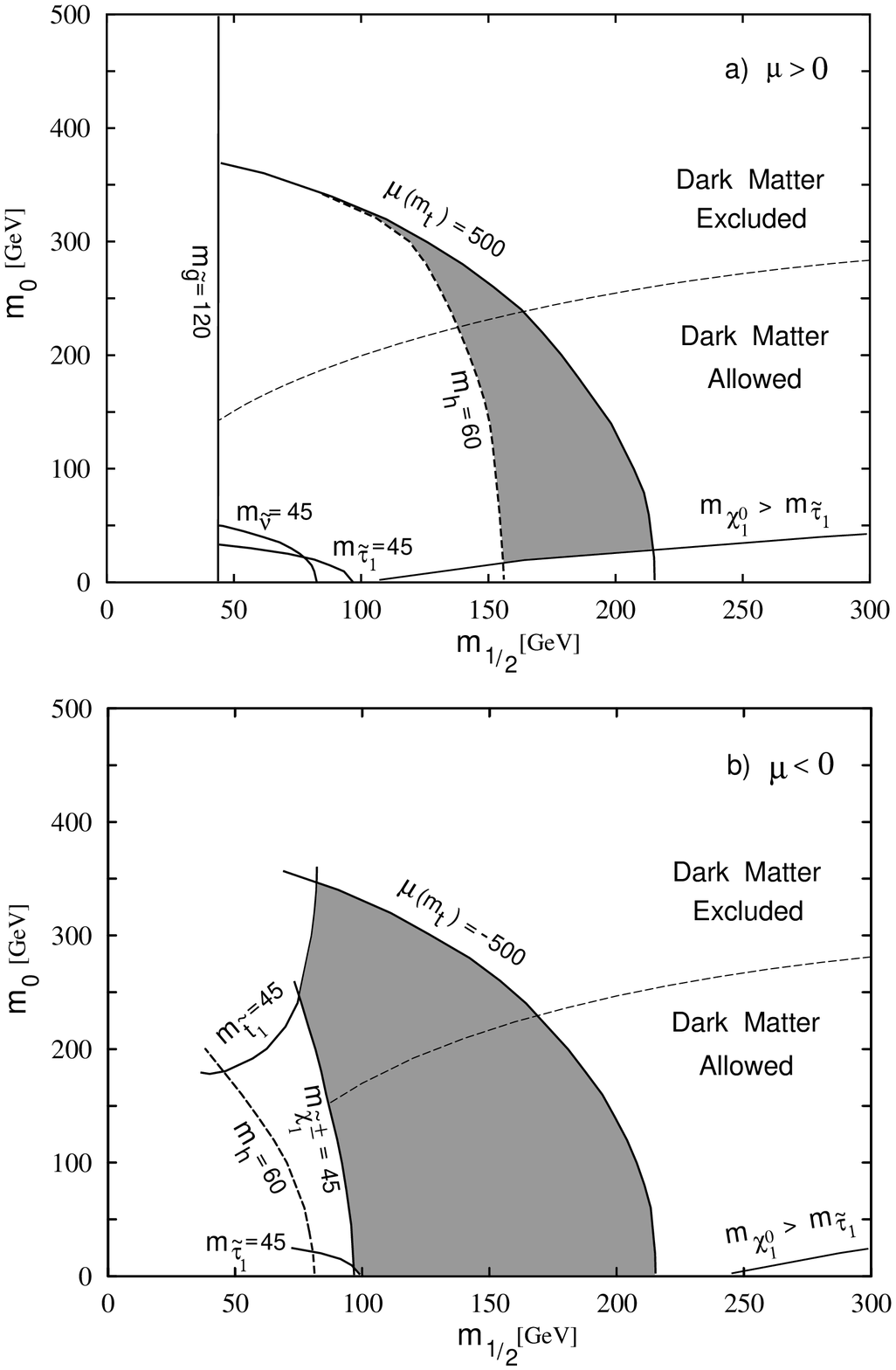}

\medskip
\parbox{5.5in}{\small Fig.~4. Allowed regions of parameter space for $m_t(m_t)
= 160$ GeV, $\tan \beta=1.47$ (a low-$\tan \beta$ $\lambda_t$ fixed-point
solution)
(from Ref.~\cite{madph801}).}
\end{center}

Hence there are solutions in the low-$\tan \beta$ $\lambda_t$ fixed-point
region which satisfy EWSB constraints
(as well as our naturalness criterion) at the one-loop level.
Note that the $\mu < 0$ solutions have more allowed parameter space
than do the $\mu > 0$ solutions. A few additional remarks are
pertinent:
the prediction for $m_h$ in the low-$\tan \beta $ region is particularly
sensitive to higher order corrections\cite{lp3},\cite{kys},\cite{hempho}.
Hence the precise location of the
$m_h=60$ GeV contour is somewhat uncertain.
Also, the dark matter line in Figure 4
should be regarded as semi-quantitative only since the contributions of
$s$-channel poles that can enhance the annihilation rate have been neglected
\cite{dn}.

\section{Conclusions}

Given only two reasonable assumptions
\begin{itemize}
\item unification of couplings and
$\lambda_b(M_G) = \lambda_{\tau}(M_G)$  at the GUT scale.

\item $m_b(m_b) = 4.25 \pm 0.10$ GeV
\end{itemize}
we are restricted to the fixed-point region of $m_t - \tan\beta$
parameter space. With only one additional assumption,
$m_t(m_t) \ltap 175$ GeV, we are restricted to either

   1) $\tan\beta \ltap 2$ ($\lambda_t$ fixed point), or

   2) $\tan\beta \gtap 50$ ($\lambda_b = \lambda_{\tau}$ fixed point).
The small $\tan\beta$ solution is favored by proton decay and
flavor changing neutral current constraints.
We investigated the additional constraint imposed by radiative electroweak
symmetry breaking upon this first region, and found solutions
which are both experimentally viable and meet the
naturalness criterion
$|\mu(M_Z)| \simeq |\mu(m_t)| < 500$ GeV.

\section {Acknowledgements}
We would like to thank Roger Phillips for collaboration.
This research was supported
in part by the University of Wisconsin Research Committee with funds granted by
the Wisconsin Alumni Research Foundation, in part by the U.S.~Department of
Energy under contract no.~DE-AC02-76ER00881, and in part by the Texas National
Laboratory Research Commission under grant nos.~RGFY93-221 and FCFY9302.
MSB was supported in
part by an SSC Fellowship. PO was supported in
part by an NSF Graduate Fellowship.

\end{document}